\documentclass[twocolumn,nofootinbib,amsmath,amssymb,aps,prd,balancelastpage,superscriptaddress]{revtex4-1}

\usepackage{color}
\usepackage[active]{srcltx}
\usepackage{amsmath,amsfonts,amssymb,amsthm,amstext,amscd,eucal,srcltx}
\usepackage{epsfig,graphicx,bm}
\usepackage{epstopdf, epsf}
\usepackage{dcolumn}
\usepackage{hyperref}

\newcommand{\be}{\begin{equation}}
\newcommand{\ee}{\end{equation}}

\newcommand{\bse}{\begin{subequations}}
\newcommand{\ese}{\end{subequations}}
\newcommand{\bea}{\begin{eqnarray}}
\newcommand{\eea}{\end{eqnarray}}
\newcommand{\ba}{\begin{array}}
\newcommand{\ea}{\end{array}}
\newcommand{\bc}{\begin{center}}
\newcommand{\ec}{\end{center}}

\begin{document}
\preprint{IPM/P-2012/009}  
\vspace*{3mm}

\title{$\mathcal{H}$olographic $\mathcal{N}$aturalness and $\mathcal{T}$opological $\mathcal{P}$hase $\mathcal{T}$ransitions}%

\author{Andrea Addazi}
\email{andrea.addazi@lngs.infn.it}
\affiliation{Center for Theoretical Physics, College of Physics Science and Technology, Sichuan University, 610065 Chengdu, China}
\affiliation{INFN sezione Roma {\it Tor Vergata}, I-00133 Rome, Italy}

\begin{abstract}
\noindent

We show that our Universe lives in a topological and non-perturbative vacuum state full of a large amount of hidden quantum hairs, the {\it hairons}.
We will discuss and elaborate on theoretical evidences that the quantum hairs are related to the gravitational topological winding number in {\it vacuo}. 
Thus, hairons are originated from topological degrees of freedom, holographically stored in the de Sitter area. 
The hierarchy of the Planck scale over the Cosmological Constant (CC) is understood as an effect of a Topological Memory intrinsically stored in the space-time geometry.
Any UV quantum destabilizations of the CC are re-interpreted as Topological Phase Transitions, related to the desapparence of a large ensamble of
topological hairs. This process is entropically suppressed, as a tunneling probability from the N- to the 0-states. Therefore, the tiny CC in our Universe is a manifestation of the rich topological structure 
of the space-time. 
In this portrait, a tiny neutrino mass can be generated by quantum gravity anomalies and accommodated into a large N-vacuum state.
We will re-interpret the CC stabilization from the point of view of Topological Quantum Computing. 
An exponential degeneracy of topological hairs non-locally protects the space-time memory from quantum fluctuations 
as in Topological Quantum Computers.

\end{abstract}

\maketitle


{\it Introduction: topological aspects of the cosmological stabilization}.  In our recent works, we elaborated on the holographic stabilization 
of the Cosmological Constant (CC) from the large entropy content {\it in vacuo}.
The CC problem is re-thought considering a holographic decoherence 
induced by a large ensamble of dynamical quantum hairs, dubbed hairons \cite{Addazi:2020hrs,Addazi:2020vhq}\footnote{See also Ref.\cite{Addazi:2016jfq} for further considerations on the relation among quantum gravity and the CC}.
All CC quantum instabilities are suppressed as 
$$e^{-S}\sim e^{-N}\, ,$$
where $S$ is the Universe entropy and $N$ is the number of hairons. 
We will elaborate on this picture and on the multitude of inter-related questions:
{\it what are the hairons? How are they originated from?} {\it What is the mechanism behind the quantum information storage and memory in space-time?}

In this paper, we explore a curious and potentially insightful relation of ${\bf the \, N}$ with the gravitational instantons
and the quantum gravity topological sector. We show that the entropic solution of the CC problem 
is also related to a {\it topological protection} against CC-Planckian mixings. 
Intriguingly, this corresponds to the Topological Phase Transitions
interpolating vacuum states with different gravitational topological winding numbers.

Let us consider the topological sector of gravity:
\begin{equation}
\label{Sg}
\mathcal{S}= \int \mathcal{E}_{G}\, , \
\end{equation}
where 
$$\mathcal{E}_{G}=dC_{G},\,\,\,\, C_{G}=\Gamma d\Gamma-\frac{3}{2}\Gamma \Gamma \Gamma\, , $$
\begin{equation}
\label{EdC}
\mathcal{E}_{G}=R\tilde{R}\, .
\end{equation}

The $\mathcal{E}_{G}$ is a topological term; its integral provides for the winding integer $N$:
\begin{equation}
\label{aa}
\mathcal{S}_{N}=N\, . 
\end{equation}
There is a multitude of deep relations among the topological action, the CC, the Planck scale and the winding number:
\begin{equation}
\label{quantu}
\mathcal{S}_{N}=\frac{1}{\alpha_{G}(\Lambda_{N})}=\frac{M_{Pl}^{2}}{\Lambda_{N}}\sim N\, . 
\end{equation}

An Euclidean action as Eq.\ref{quantu} also corresponds to the entropy of the thermodynamical space-time system:
\begin{equation}
\label{SNN}
S=\log\, \Omega\sim N\, ,
\end{equation}
where $\Omega$ is the configuration space.
This is leading us to a simple identification of the topological winding number and the entropic content of the space-time. 
In this sense, the space-time entropy is re-interpreted as a topological index.

Therefore, a so tiny CC, as the one observed in our Universe (around $\Lambda\sim 10^{-123}M_{Pl}^{2}$), would correspond 
to the 
$$\bar{N}\sim 10^{123}$$
vacuum eigenvalue of the $|\bar{N}\rangle$ (see \cite{Addazi:2020hrs,Addazi:2020vhq}). 

Rephrased in this way, the quantum stability issue of the CC appears 
as a problem of why the $|\bar{N}\rangle$ eigenstate would not spontaneously flow to the $|0\rangle$,
in turn corresponding to the UV divergent case as $\Lambda=M_{Pl}^2/N\rightarrow \infty$ $(N\rightarrow 0$). 
To compute the quantum bubble diagrams would be re-thought, not only 
as an integral on the momenta, but also as a divergent series of the topological winding number.
Indeed, a particle with a momenta $p$ would probe a number of hairons as $N\sim M_{Pl}^{2}/p^{2}$,
i.e. the UV momentum corresponds to a IR topological number divergence. 

Therefore, the spontaneous flow of the CC from the IR to the UV 
corresponds to the disapparence of $N$ winding numbers, costing N-entropic units:
\begin{equation}
\label{cost}
\langle 0|N\rangle\, . 
\end{equation}
This transition thermodinamically costs as an exponential suppression, i.e. as the Eq.1.
This process is also interpreted as a quantum tunneling, related to an N-instanton. 
Thus, the quantum destabilization of the CC 
would cross several orders of magnitude, rendering it probabilistically forbidden as 
\begin{equation}
\label{PPP}
\mathcal{P}(N\rightarrow 0)\sim e^{-\Lambda/M_{Pl}^{2}}\sim e^{-10^{123}}\, . 
\end{equation}

This is offering a simple re-interpretation of the CC problem.
When the theoretical physicist calculates the 
vacuum to vacuum correlator,
\begin{equation}
\label{ca}
\langle 0 |\Phi(x)\Phi(x)|0\rangle
\end{equation}
of any SM fields indicated here as $\Phi$,
he/she finds a disappointing result: quantum field theories typically predict quartic UV divergences. 
However, this correct result is typically misunderstood and misleadingly interpreted as 
an instability of the CC state. 
Indeed, a possible instability crossing from the IR to the UV domains
is passing from a non-trivial correlator 
\begin{equation}
\label{coca}
\langle N|\Phi(x)\Phi(x)|0\rangle\, .
\end{equation}
Rephrased in this way, quantum instabilities would work into the system
for re-turning from a completely disorder to a fully ordered state.
Indeed, 
\begin{equation}
\label{coca}
\langle N|\Phi(x)\Phi(x)|0\rangle=e^{-N}\langle 0|\Phi(x)\Phi(x)|0\rangle\,. 
\end{equation}
where 
$$e^{-N}=e^{-S}=\frac{1}{\Omega(N)}\, ,$$
and $\Omega(N)$ is the probability configuration volume. 

Therefore, any quantum fluctuations cannot destabilize a maximal entropic state
because of the large entropic barrier. 
The high entropy state can reach the UV domain only through an exponentially suppressed tunneling. 

On the other hand this is naturally leading to the concept of an emerging CC
as a thermal effect of a large number of hairons accumulated in vacuo:
\begin{equation}
\label{te}
T_{hairons}\sim \sqrt{\Lambda}\sim \frac{1}{\sqrt{N}}M_{Pl}\, . 
\end{equation}
It is certainly interesting that the number of hairons 
scales as the Topological winding number. 
Indeed, this is a signal that the entropic protection from CC-Planckian mixings 
is deeply related to the topological space-time properties. 
In other words, the $|0\rangle$ and the $|N\rangle$ have an inequivalent topological structure
that can never be interpolated by a trivial geometric deformation on the space-time manifold.
The $|N\rangle$ and the $|0\rangle$ can be transformed each others through a N-instanton 
with a multi-spherical topology.
In other words, the dictionary among the $N$ and the $0$-vacuum states and the instanton topology is as follows:
\begin{equation}
\label{SD}
\langle 0|N\rangle:\,\,\,\,S_{4}\rightarrow_{tunneling} \rightarrow S_{2}\times S_{2}\times .... \times S_{2}=\prod_{n=0}^N(S_{2})^{n}\, .
\end{equation}
The $0-N$ quantum transition has a precise topological and geometrical sense,
corresponding to a topological phase transition of the space-time. 
As we will see, this is related, in the real (non-euclidean) space-time,
to gravitational topological defects puncturing the space-time boundary. 
In other words, the N labels the space-time defects.
This also implies that the fundamental temperature in {\it vacuo},
providing for the CC, is emerging as a topological effect 
 from a large N of gravitational defects. 
 On the other hand, Eq.\ref{SD} corresponds to a first order phase transition of 
 the gravitational susceptibility in vacuo as 
 \begin{equation}
 \label{RRti}
\langle R\tilde{R}\rangle_{N=0}\rightarrow \langle R\tilde{R}\rangle_{N}\, .
 \end{equation}

As we will see, this fact is leading to 
several potential breakthroughs towards our understanding of space-time memory and CC stabilization.

\vspace{0.1cm}

{\it Dynamical Relaxation.} 
Let us now promote the N to a dynamical field, 
as
\begin{equation}
\label{QCAA}
N \rightarrow \varphi(x)\, , 
\end{equation}
where $\varphi(x)$ is the relaxon field. 
This corresponds to 
\begin{equation}
\label{aj}
\varphi(x)=\frac{M_{Pl}^{2}}{\Lambda(x)}\, ,
\end{equation}
where $\Lambda$ is thought as dynamical.
If $\varphi(x)$ evolves as a runaway field to the asymptotic infinite, then the CC will dynamically flow to the large-$N$-vacua.  
This mechanism is understood as Topological Phase Transitions among the different topological vacuum states
labelled by the winding number. 
The relaxation mechanism can be specialized in a cosmological set-up.
Indeed, Eq.\ref{aj} in CC corresponds to the Hubble rate of the Universe.
Considering a scalar field in cosmology, provoking the expansion of the Universe, 
it would be related to the Hubble rate as
\begin{equation}
\label{jaja}
\varphi=\frac{1}{H^{2}}=\frac{1}{\frac{1}{2}\dot{\phi}+V(\phi)}\, ,
\end{equation}
where the other $\phi$ field is a Dynamical DE scalaron.
If the potential drives the scalar field $\phi$ to zero, then $\varphi$ will run-away to a an asymptotic attractor point. 
This is interpreted as the dynamical generation of N-quanta, topologically stored in every Planckian volume 
and, therefore, appearing out with the space-time expansion. 
This process would be spontaneous since related to the entropic attractor 
\begin{equation}
\label{SD}
S=\varphi(t)=\frac{M_{Pl}^{2}}{\Lambda(t)}\, ,
\end{equation}
In other words, scenarios where the $\phi$ tends to relax to zero are probabilistically favored.
Indeed, for maximizing the entropy,
the $\varphi$ field increases without any upper bound. 
Such a phenomena is interpreted as a dynamical proliferation in time of the number of hairons,
corresponding to the CC screening effect \footnote{An alternative axion-inspired approach can
be found in Ref.\cite{Alexander:2018djy}. Another self-adaptive holographic mechanism was suggested in Ref.\cite{Charmousis:2017rof}. }. 

What is the topological meaning of such a relaxation mechanism?
 The N winding number was promoted to a dynamical field;
the fact that the $\varphi$ increases up to infinity is interpreted 
as a spontaneous increasing of the topological complexity of the space-time.
Such a phenomena corresponds to a spontaneous cascade of topological transmutations.
This process asymptotically tends to accumulate an infinite number of gravitational defects in space-time.
In other words, the $\varphi$ can be interpreted as a topological order mean field.

\vspace{0.1cm}

{\it Dark Energy and Neutrino mass.} The Cosmological Relaxation mechanism 
can be related to the generation of neutrino masses from gravitational anomalies \cite{Dvali:2016uhn}. 

Indeed gravitational instantons induce anomalous terms that can dynamically break the $U_{A}(1)$ 
axial symmetry, generating a neutrino mass term from the neutrino current anomaly:
\begin{equation}
\label{neutrino}
\partial_{\mu}J^{\mu}_{5}=\mathcal{E}_{G}\, .
\end{equation}
If the vacuum state would be empty of the N-quanta, 
then, paradoxically, the anomalous term 
would be divergent 
\begin{equation}
\label{nam}
\langle 0|\, \mathcal{E}_{G}\, |0\rangle \sim {\rm lim}_{N\rightarrow 0} \Lambda_{N}\rightarrow \infty\, . 
\end{equation}

Reversing this argument, this may be a hint that we do not live in a trivial vacuum state, otherwise the neutrino would be affected 
by a new hierarchy problem. 
Fortunately, this is exactly the same issue just solved above for the CC.
Therefore, in N-state of our Universe, corresponding to the observed Hubble rate, the lightest neutrino mass is 
\begin{equation}
\label{nas}
m_{\nu}^{4}\sim G_{N}^{-1}\Lambda_{N}\sim \frac{1}{N}M_{Pl}^{4}\sim 10^{-123}M_{Pl}^{4}\sim (1\, {\rm meV})^{4}\, . 
\end{equation}

In this sense, also the neutrino mass hierarchy is understood as a manifestation of the topological memory storage in space-time.
On the other hand, the dynamical DE mechanism would relate the CC flow to zero to a dynamical neutrino mass. 
This also means that the neutrino mass is generated and stabilized as an environmental and thermal effect 
in the quantum criticality point.

\vspace{0.1cm}

{\it Coherent state portrait.} It exists a general duality among instantons and solitons (see for example Refs.\cite{Addazi:2016yre}).
The gravitational vacuum state storing hairons can be interpreted as a gravitational solitonic state, 
while the corresponding gravitational instantons as a tunneling of a soliton from nothing. 

The soliton can be viewed as a coherent state of $N$-hairons,
in turn related to the topological winding number. 
The solitonic quantum state can be expressed as a tensor product of coherent states \cite{Dvali:2015jxa,Addazi:2018ivg}: 

\begin{equation}
\label{sols}
|S\rangle=\prod_{\otimes k}|\alpha_{k}\rangle
\end{equation}
where
\begin{equation}
\label{aal}
|\alpha_{k}\rangle=e^{-\frac{1}{2}|\alpha_{k}|^{2}}\sum_{n_{k}=0}^{\infty}\frac{\alpha_{k}^{n_{k}}}{\sqrt{n_{k}!}}|n_{k}\rangle\, .
\end{equation}
where $|n_{k}\rangle$ are number eigenstates of 
hairons with momentum $k$\, . 

The mean occupation number is defined as 
\begin{equation}
\label{maiam}
N=\int_{k}N_{k}=\int_{k}\alpha_{k}^{*}\alpha_{k}\, , 
\end{equation}
where 
\begin{equation}
\label{lal}
\hat{N}_{k}=\hat{a}_{k}a_{k}\rightarrow N_{k}=\langle S|\hat{a}_{k}^{\dagger}a_{k}|S\rangle\, . 
\end{equation}

An intrinsic energy scale of the soliton can be obtained
as
\begin{equation}
\label{modes}
E=\int_{k}|k|N_{k}\, . 
\end{equation}

Considering any hairons as Planckian, a Universe-size soliton 
corresponds to 
\begin{equation}
\label{planck}
E=\sqrt{N}M_{Pl}\, .
\end{equation}

Within this picture, the enucleation of the hairon solitonic state \footnote{To be more accurate, there would also be a factor $1/2$ in the exponential, essentially not important in our discussions.  }
is 
\begin{equation}
\label{hairo}
\langle 0|S \rangle=e^{-N}\, .
\end{equation}

This transition amplitude is related to the solitonic synthesis from nothing, 
in turn dual to a gravitational instanton solution. 
Thus, the instantonic language can be rephrased in terms of the coherent state formalism. 

\vspace{0.1cm}

{\it Cosmological Phase Transitions and Criticality.} Now, we move on the analogy among 
gravitational solitons with other known critical systems. 
As remarked by Kibble and Zurek, 
topological defects can appear in the early Universe
or in condensed matter system around a critical temperature 
$T_{c}$ \cite{Kibble1,Kibble2,Zurek1,Zurek2,Zurek3}.
An important example of this phenomena 
is the notorious Berezinskii-Kosterlitz-Thouless transition (BKT transition) \cite{Kosterlitz:1973xp}.
It is related to a phase transition in the two dimensional XY models,
in turn well describing a certain class of 2D systems 
in condensed matter -- including a large class of 
thin disordered superconductors. 
Around a critical temperature $T_{c}$, 
the BKT transition is related to the 
the transition from a pair 
to an unpaired couple 
of vortices and anti-vortices. 
Indeed, in the 2D XY models,
the (anti)vortices are 
topologically stable structures. 
The production of vortices becomes thermodynamically 
efficient around the $T_{c}$. 
In other words, for $T<T_{c}$,
the bounded vortex-antivortex system 
has a lower energy and entropy than 
the two decouple ones. 

\begin{figure}[ht]
\centerline{ \includegraphics [width=1\columnwidth]{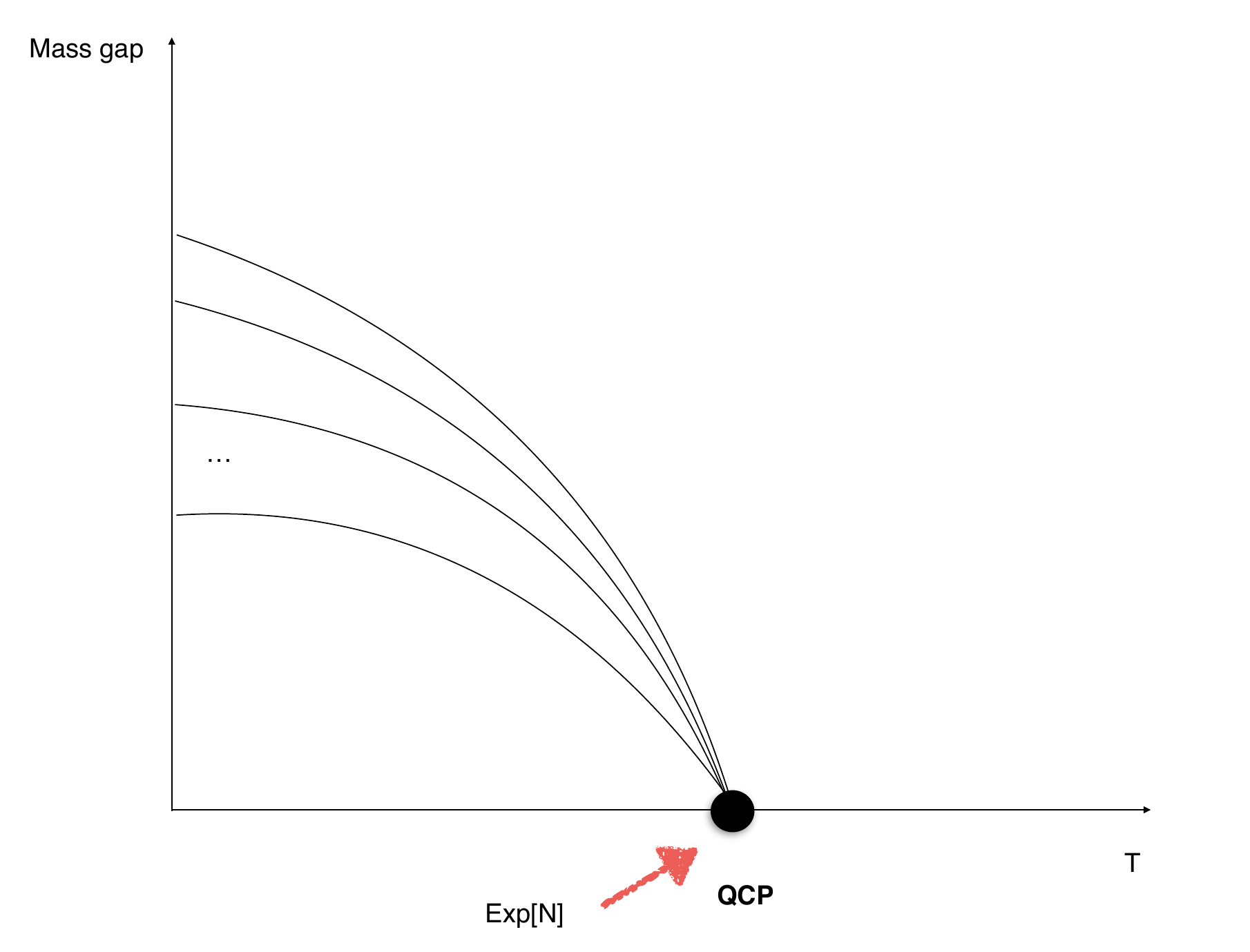}}
\caption{The de Sitter space-time lives on a topological quantum critical point with an exponential degeneracy. The mass gap among the N-states flows to zero around the critical temperature $T_{c}\sim \sqrt{\Lambda}$,
where $\Lambda$ is the CC. The degeneracy corresponds to the hairon number $N$ on the same ground state. }
\end{figure}

The critical temperature is understood as 
the saddle solution of the free-energy 
as 
\begin{equation}
\label{FF}
F=E-TS\, \rightarrow T_{c}\, . 
\end{equation}

Now, we will rise a similar question in the {\it arena }
of quantum gravity. Is there any critical temperature 
related to a topological phase transition?

We will see that the answer is {\bf yes}
and it is exactly related to the Hawking temperature. 
Indeed, the minimal possible $F$ for a thermodynamical system 
with a fixed internal energy -- in the BH case its own mass --
corresponds to the maximal entropy. 
In turn, the entropy is maximized if all the information is re-organized in a BH holographic state. 
The topological defects mentioned above are here replaced by gravitational solitons and instantons. 

Indeed, in the case of BHs and de Sitter,
the partition function is dominated by the 
euclidean action 
\begin{equation}
\label{N}
\mathcal{S}_{N}=\beta \bar{E}=\frac{\beta^{2}}{L_{Pl}^{2}}=N
\end{equation}
 as follows
\begin{equation}
\label{ZZ}
Z=\sum{\rm exp}(-\beta E)\simeq {\exp}(-N+O(1/\sqrt{N}))\, . 
\end{equation}

The free-energy is extremized on the saddle solution Eq.\ref{N},
corresponding to 
\begin{equation}
\label{FF}
F=\langle E \rangle_{BH}/2= T_{c} S\sim \beta_{c}\sim \sqrt{N}L_{Pl}\, ,
\end{equation}
where $\langle E\rangle \sim M_{BH}$ is the BH internal energy, corresponding to the BH mass
(here we omit the numerical pre-factors, as inessential for our discussions while $G_{N}=1$).
This is related to the Holographic entropy
and temperature of the de Sitter space-time. 
This means that a Universe on a de Sitter phase lives on a state 
of criticality.
On the other hand, Eq.\ref{FF} does not correspond to a global minima,
the F derivative is negative on the critical temperature
\begin{equation}
\label{F}
\frac{dF}{dT_{c}}=-\frac{1}{T^{2}_{c}}. 
\end{equation}

The fact that both the free-energy and its derivative diverge for $T_{c}\rightarrow 0$
can be interpreted as the evidence that the zero CC
corresponds to a state of super quantum criticality and a first order phase transition.
Indeed, Eq.\ref{F} is also related to the divergence of the 
heat capacity of de Sitter space-time.

In analogy with the phase transition theory 
and critical phenomena,
one would imagine that this corresponds to a divergence 
of the correlation length 
\begin{equation}
\label{ll}
\zeta\sim (T-T_{c})^{-\nu}, 
\end{equation}
where $\nu$ is a critical exponent.
This is certainly self-consistent with the (A)dS/CFT approach,
as the system is around a conformal critical point. 
We will see later that this fact has a precise topological entanglement entropy interpretation. 
Such a phenomena is related to the emergence of topological order
 around the critical temperature $T_{c}$.
From the macroscopic space-time portrait, 
this corresponds to a high degeneracy of the ground state.
Microscopically, this  is related to
 a long-range quantum entanglement. The Black Hole is viewed as a highly degenerate and fully entangled N-state
 as typically happening in topological systems \footnote{Similar ideas were proposed in the Gravitational Bose-Einstein condensate approach, where the degeneracy is understood as 
a large number of gapless Bogoliubov modes \cite{Dvali:2012rt,Dvali:2013vxa,Dvali:2012en}. It is certainly possible that our approach would be a dual portrait capturing the same quantum critical phenomena from the topological prospective.  }. 

In the sense of topological phase transitions,
the entropy assumes the role of 
a topological index
of the space-time complexity:
the entropy is related to the space-time Euler characteristic 
or the genus, in turn related to the topological winding number. 
Indeed, the entropy phase transition is intimately 
related to the jump of the gravitational susceptibility 
$$\langle R\tilde{R}\rangle_{N=0}\rightarrow \langle R\tilde{R}\rangle_{M=N}\, .$$
How we will see in next sections, this fact has an interpretation 
in a BH portrait from topological quantum computing. 

In the case of de Sitter, 
a dynamical system with the $\varphi\rightarrow \infty$
would correspond to a walking critical temperature 
as $T_{c}\rightarrow 0$. 
When the criticality is reached, 
this can dynamically walk to increase the degeneracy 
in vacuo.

\begin{figure}[ht]
\centerline{ \includegraphics [width=0.8\columnwidth]{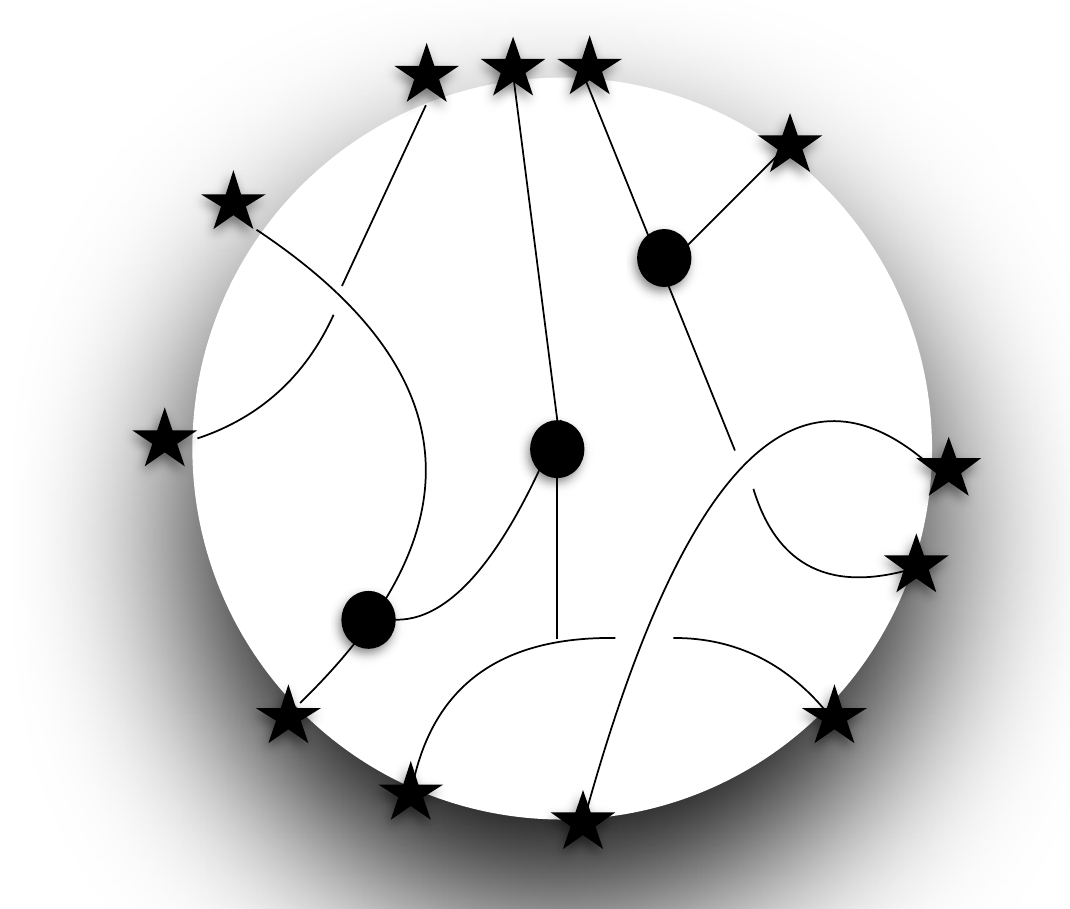}}
\caption{A space-like section of the $dS_{3}$ bulk is shown. Gravitational Wilson lines or Semiclassical Wormholes intersect the horizon providing topological hairs (black stars). They can have braids or fusions (black dots). }
\end{figure}

From the scattering processes,  the BH topological phase transition is dynamically reached from 
localizing a transplanckian CM energy into an impact parameter 
of $b\leq R_{S}=2G_{N}\, E_{CM}\sim T_{c}^{-1}$, where $R_{S}$ is the Schwarzschild radius (here we preferred to re-insert $G_{N}$). 
Aspects of scatterings are considered in another paper in preparation.

\vspace{0.2cm}

{\it Hairons as Punctures on the Horizon.} We now elaborate on the idea that 
holographic hairons correspond to bulk defects. 
A possibility, inspired by the Loop Quantum gravity approach 
(see Refs.\cite{Rovelli:1996dv,Ashtekar:1997yu}) is that
hairons are topological punctures on the BH or de Sitter horizons
generated by Gravitational Wilson lines in the Bulk geometry.
From the semiclassical approach, conversely, this can be related to 
gravitational quantum tunnels puncturing the BH horizon \cite{Addazi:2017xur,Chen:2018aij,Addazi:2019vch}.
A third vision of the phenomena is in terms of gravitational coherent states ramifying in the bulk 
by means of the duality among instantons and coherent wave functions. 
  
  The action would correspond to 
a topological sector coupled with the puncture contribution:


\begin{equation}
\label{puntu}
N=\mathcal{S}_{E}=\int_{\Delta/\{\prod_{p=1}^{N}\{p\} \}}(\Gamma d\Gamma+\frac{2}{3}\Gamma^{2})\,.
\end{equation}
where $\Delta$ is the continuous space-time Horizon and $\prod_{p}\{p\}$ is the ensamble of points puncturing the horizon.

Now, an interesting question is as follows.
Let us suppose to have a quantum state of hairons 
as 
\begin{equation}
\label{state}
|h_{1},....,h_{N}\rangle\, .
\end{equation}

What is the hairon spin-statistics? Are they bosons or fermions?
What happens to the Eq.\ref{state} under permutational operations acting on hairon fields?

Naively one would think that, since we are in $3+1$ dimension, 
the permutation of a couple of hairons would just give
a $\pm 1$ corresponding to boson or fermion statistics.
However, we will show a simple argument to convince the reader that
this is not what is happening in our case: the hairons have an {\it anyon statistics} as happening in 
2D topological materials
\cite{LM,Goldin,W,Wu:1984hj,Wu,Haldane}  \footnote{It is possible that such an effect may be testable considering topological scatterings of Standard Model particles on space-time anyons. These effects
may percolate into effective tiny violations of the Spin Statistics in the SM sector, testable in underground experiments with high precision   \cite{Addazi:2017bbg,Addazi:2018ioz,Addazi:2019ruk}.}.

To fix our idea, let us start from comparing the hairon picture with what happens to the case of two normal point-like particles 
and hairons. 
In 3D, one path $\gamma_{2}$, encircling the first particle 
can be always continuously deformable to another path $\gamma_{1}$
that does not encircle the second particle.
In other words, the path can be deformed in such a way to pass just behind the second particle.
On the other hand, the $\gamma_1$ loop can be contracted to just a point.
This corresponds to the condition, on the wave function of the system, as
\begin{equation}
\label{con}
|\psi(\gamma_{2})\rangle=|\psi(\gamma_{1})\rangle=|\psi(0)\rangle\, . 
\end{equation}
Eq.\ref{con} means to relate the wave function of the double circling path
to the one with no any circling as 
\begin{equation}
\label{caa}
|\psi(\gamma_{2})\rangle=R^{2}|\psi(0)\rangle,\,\,\, R^{2}=1\, .
\end{equation}
Thus, we can have only to cases, corresponding to bosons and fermions: 
$$R_{B=+1,F=-1}=\pm 1\, .$$
However, for hairons, this is expected to be not true:
a path circulating around one hairon 
cannot be contracted without crossing the gravitational Wilson lines 
or the gravitational instantons having the hairon punctures as edge states. 
Therefore, in our case, $R$ would have a more general form \footnote{Let us remark that 
$R$ may also be no-unitary, as well as $T$ non-hermitian if we consider non-abelian non-unitary groups as $SO(3,1)$ or $SU(1,1)$} as
\begin{equation}
\label{ge}
R=e^{iT}\, . 
\end{equation}
In general, the $R$ transformation can be related to abelian or non-abelian 
structures, in turn corresponding to $T$ as just a phase number or a matrix. 
In the case of quantum gravity, we would expect a non-abelian nature of anyonic
hairons, for two motivations: i) from the algebraic prospective, gravity is more similar to non-abelian Yang-Mills theories 
rather than QED: if reformulated as a gauge theory, it would correspond to 
a $SO(3,1)$ local group; ii) non-abelian anyons are related to an exponential degeneracy 
of quanta populating the lowest energy level, compatible with the criticality condition 
envisaged in the case of Black Holes as well as for their exponentially enhanced memory storage
$e^{N}$. 
In a de Sitter space-time,
the energy levels (to not be confused with the CC levels) correspond to 
\begin{equation}
\label{ka}
E_{N}^{2}=NM_{Pl}^{2}
\end{equation}
with a degeneracy of 
$$S_{N}\sim e^{N}\, . $$
The next level is at $N-1$,
therefore the level splitting 
is 
\begin{equation}
\label{DeltaNN}
\Delta_{N-N-1}^{2}=M_{Pl}^{2}\, . 
\end{equation}
The degeneracy of any next level is scaling as the N-number. 
By the way, this also offers a Fermi Golden Rule explanation of why 
transitions from the N-level to the zero one are exponentially disfavored 
as dressed by the density state factor 
\begin{equation}
\label{Fermi}
\rho_{1}/\rho_{N}\sim e^{-N}\, . 
\end{equation}

It is worth to remark that the $R$-transformation does not change the ground state Eq.\ref{ka}
of the system, compatible with the exponential degeneracy.

Elaborating on the duality among gravitational instantons and gravitational Wilson lines,
puncturing the horizon, we wish to propose a possible vision of how black holes as well as the de Sitter Universe
may store quantum informations. The punctures on the Horizon can correspond
to several different world-lines. As we said before, punctures may be re-interpreted as
our hairons. The puncture corresponds to a sort of gravitational defect in the bulk geometry,
very much in analogy with the Ahranov-Bohm effect. 
Therefore, the BH can elaborate and store information in a large number of possibilties
provided by the different {\it braids} on the hairon world-lines.
This picture is inspired by topological quantum computers,
tipically storing qu-bits with several different possibilities of braiding the anyons (see Ref.\cite{Nayak:2008zza}
 for a complete review on these subjects). 
The idea is that information infalling inside a black hole is computed as a series  
of braidings. Hairons are non-abelian anyons 
that obey to non-Abelian braiding statistics. 
In space-time, Quantum information is stored in states with multiple hairons, which have a topological degeneracy and braids.

Let us consider the system of $N$ non-abelian anyons in the 
ground degenerate state:
$|\Psi_{n}(z_{1},z_{2},...,z_{N})\rangle\,,$
where $z_{j}$ are the hairon coordinates
and $n=1,...,D$ labels for a D-dimensional protected subspace. 
We can consider the $\gamma$-path around the $z_{j}$,
winding one anyon around another.
Let us consider the following transformation on the hairon state:
\begin{equation}
\label{evo}
|\Psi_{n}(z_{1},z_{2},...,z_{N})\rangle\rightarrow \sum_{m=1}^{D}W_{nm}(\gamma)|\Psi(z_{1},z_{2},...,z_{M})\rangle\, , 
\end{equation}
where
\begin{equation}
\label{kakp}
W(\gamma)={\bf P}\, {\rm exp} \oint_{\gamma}{\bf \Gamma} \cdot d{\bf z}\, ,
\end{equation}
where ${\bf P}$ represents the path ordering. $W$ is nothing but a Gravitational Wilson line. 

The $W$ transformation can be viewed as 
\begin{equation}
\label{WW}
W=F^{-1}R^{2}F\, ,
\end{equation}
where $F$ is describing the Fusion among 
hairons into a new hairon.
At this point of the discussion, this equivalence may be not clear, but 
we wish to shine light on these aspects in the following.
In our portrait, the hairon evolutions are limited by these
three rules: i) they can be created or annihilated, in pairwise fashion;
ii) hairons can be fused to generate other hairons;
iii) hairons can be exchanged as in Eq.\ref{evo}.
The F(usion) in a collective non-local  property of the hairons,
as intrinsically non-local is the nature of the Wilson-lines. 
 
\begin{figure}[ht]
\centerline{ \includegraphics [width=1.0\columnwidth]{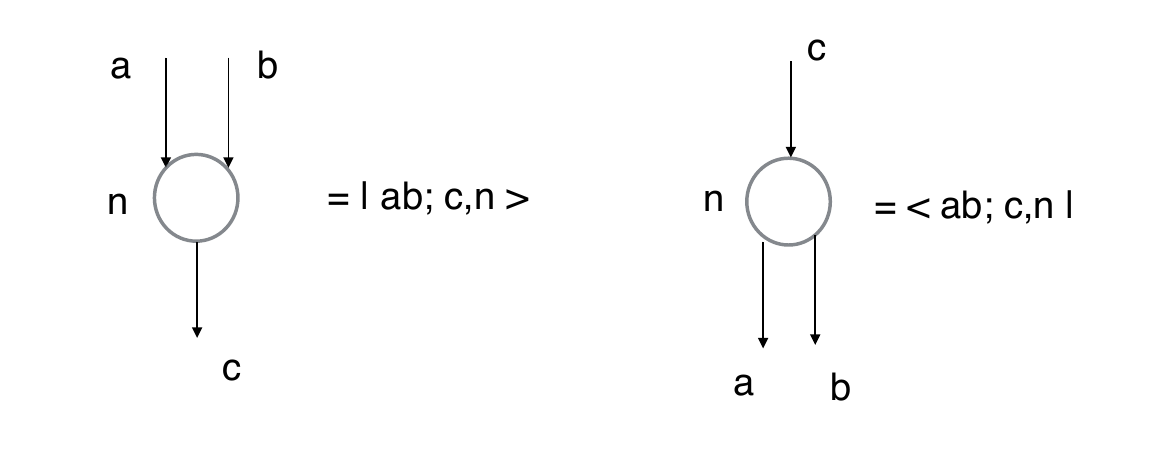}}
\caption{A diagramatic representation of the anyon states.}
\end{figure}
\begin{figure}[ht]
\centerline{ \includegraphics [width=0.7\columnwidth]{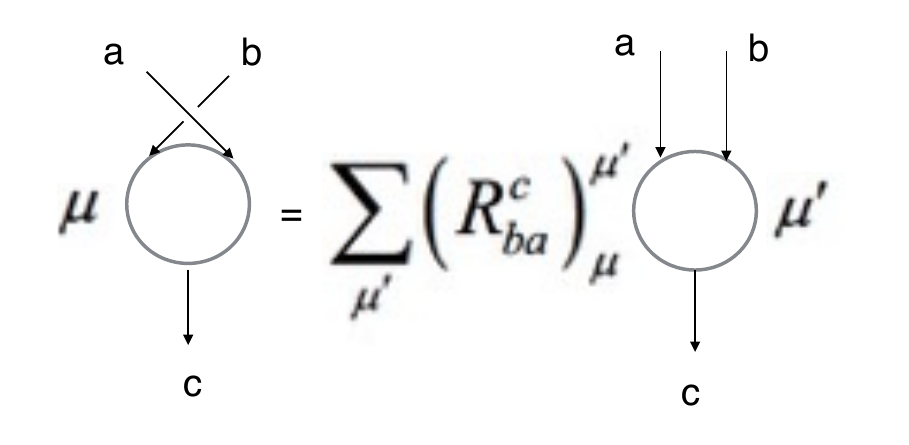}}
\caption{A diagrammatic representation of the F(usion) operation is displayed.}
\end{figure}
\begin{figure}[ht]
\centerline{ \includegraphics [width=1\columnwidth]{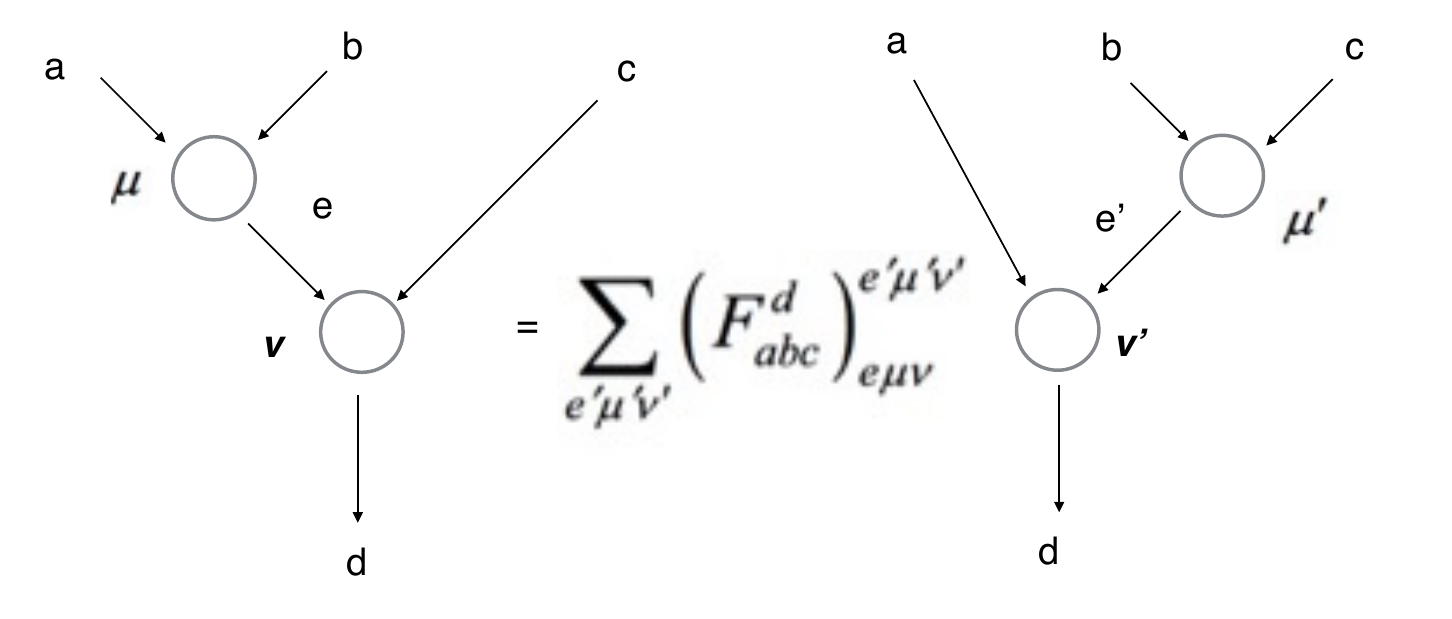}}
\caption{A diagrammatic representation of the R operation is displayed.}
\end{figure}

In the following, we will elaborate on the topological quantum operations.  
 
First of all, the hairons, or gravitational anyons, 
can interact to form a new one; the non-abelian interaction 
of them is related to a non-abelian product as 
\begin{equation}
\label{THS}
a \otimes b=\sum_{c}N_{ab}^{c}c
\end{equation}
 where
\begin{equation}
\label{aba}
a\otimes b=b\otimes a,\,\,\, a\otimes 1=a\, . 
\end{equation}
The fusion processes correspond to a vector space 
denoted as $N_{ab}^{c}$, dubbed topological Hilbert spaces (see Fig.3).
A complete orthogonal basis describing the anyon system reads as 
\begin{equation}
\label{basi}
\{|a,b,c,\mu\rangle;\, \mu=1,...,N_{ab}^{c}\}
\end{equation}

The R and F transformations acting on the hairon diagrams 
have simple pictorial representations
as in Fig. 4-5.
The two diagrams correspond to the 
state transitions:
\begin{equation}
\label{t1}
|(ab)c;ec;d\rangle=\sum_{f}(F_{abc}^{d})_{ef}|a(bc);af;d\rangle\, , 
\end{equation}
\begin{equation}
\label{t2}
|(ba)c;ec;d\rangle=\sum R_{ab}^{f}\delta_{e,f}|(ab)c;ec;d\rangle\, ,
\end{equation}
where $\delta$ is the Kronecker delta function
and f spans all possible out-results of $a,b$.

This is suggesting an information computation 
through hairon/anyon exchanges. 
A topological computation can be 
initiated as a sequence of hairons 
and then, performing a sequence of 
particle exchanges, the hairon worldlines will trace out
braids! In these processes also fusions are possible
and in principle the final state can end with a complete 
annihilation process, but, indeed, with an entropic cost. 

Such a computational system can also be dually re-interpreted as
a neural network computation in the bulk,
where hairons are the neural centers
interconnected by gravitational synapsis extending on the bulk geometry (see Ref.\cite{Dvali:2018vvx} for recent attempts 
on similar research directions).

The computational operations of the gravitational anyons 
form a braid group related to the N-punctures on the horizon as $\mathcal{B}_{N}$ (See for example Ref.\cite{Braids}).
As we mentioned, any braid is related to F and R operations as 
\begin{equation}
\label{B}
B=F^{-1}RF\, \in \mathcal{B}_{N}
\end{equation}
The generators of $\mathcal{B}_{N}$ can be viewed
as clockwise interchanges of the $i$th with 
the $i+1$th lines. 
Let us denote this generator as $\sigma_{i}$. 
The inverse operation $\sigma_{i}^{-1}$
corresponds to a counter-clockwise
rotation of $i$th and $i+1$th. 
The generators satisfy the following algebraic conditions:
\begin{equation}
\label{sig}
\sigma_{i}\sigma_{j}\sigma_{i}=\sigma_{j}\sigma_{i}\sigma_{j},\,\,\,\, {\rm if}\,\,|i-j|=1\, ,
\end{equation}
\begin{equation}
\label{sig2}
\sigma_{i}\sigma_{j}=\sigma_{j}\sigma_{i},\,\,\,\, {\rm if}\,\,|i-j|>1\,.
\end{equation}
These relations are nothing but related to the notorious Yang-Baxter equations. 

This can be defined as 
\begin{equation}
\label{Bn}
\mathcal{B}_{N}=\{ \sigma_{1},....,\sigma_{N}|\sigma_{i}\sigma_{i+1}\sigma_{i}=\sigma_{i+1}\sigma_{i}\sigma_{i+1},\sigma_{i}\sigma_{j}=\sigma_{j}\sigma_{i}\}\, .
\end{equation}

The Braids provide for Logical gates as a basis for space-time quantum computation. 

The abelianization of this group is a homomorphism mapping every $\sigma_{i}$ to 1:
\begin{equation}
\label{jaj}
h:\, \mathcal{B}_{N}\rightarrow \mathbb{Z}
\end{equation}
which also related with the winding number, in turn contributing to  the space-time entropy. 

Now, we wish to visualize the BH topological computation 
in terms of the quantum computing language.
In doing it, we need to identify the information computing in terms of 
the ($|0\rangle$,$|1\rangle$) qu-bit basis. 

Motivated by the multi-instantonic picture,
we consider pairs of gravitational flux and anti-flux units
as $|\alpha,\alpha^{-1}\rangle$. 
This can be visualized as the two edge points of a Wilson line 
puncturing the horizon. 
Let us consider two fluxon pairs $|\alpha,\alpha^{-1}\rangle$
and $|\beta,\beta^{-1}\rangle$.
We can realize a basic gate by winding counterclockwise 
the $|\alpha,\alpha^{-1}\rangle$ around the  $|\beta,\beta^{-1}\rangle$.
This operation transforms the first fluxon state as 
\begin{equation}
\label{aam}
|\alpha,\alpha^{-1}\rangle\rightarrow |\beta \alpha \beta^{-1},\beta \alpha^{-1}\beta^{-1}\rangle \, . 
\end{equation}
By means of this definition, we can identify a computational basis as 
follows:
\begin{equation}
\label{zeun}
|0\rangle=|\alpha,\alpha^{-1}\rangle,\,\,\, |1\rangle=|\beta \alpha \beta^{-1},\beta \alpha^{-1}\beta^{-1}\rangle\, . 
\end{equation}

This is not the only possible basis choice, but it is certainly a simple one. 
Indeed the single isolated $\alpha$ appears as identical to the 
$\beta \alpha \beta^{-1}$ as non-locally next to each others.
In this case, the states in Eq.\ref{zeun} do not have any quantum superposition.
In other words, quantum memory is protected by decoherence 
induced by the environment since non-locally stored 
as a phase effect induced by the fluxon-antifluxon entanglement. 

\begin{figure}[ht]
\centerline{ \includegraphics [width=1.0\columnwidth]{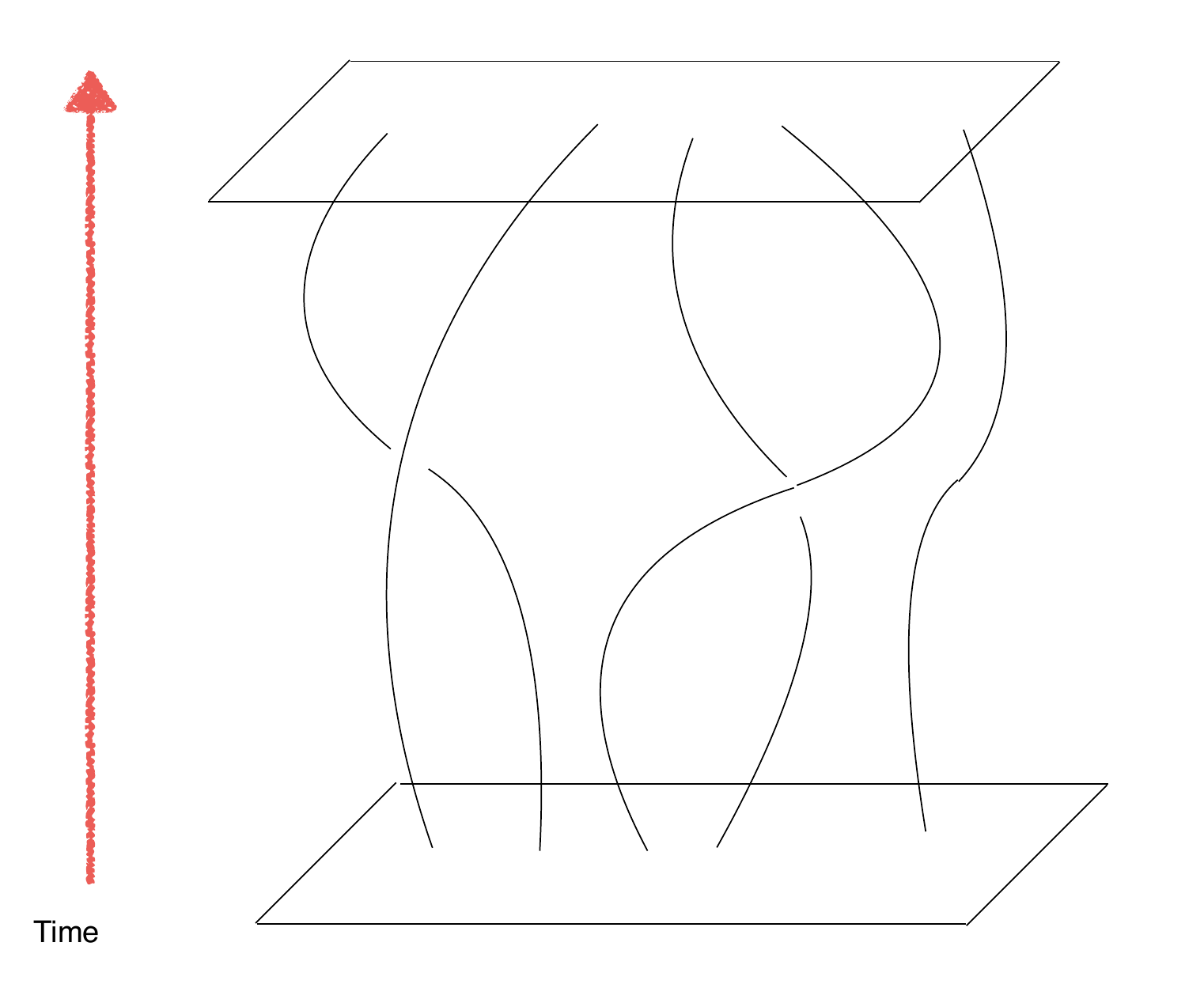}}
\caption{The topological quantum computing of space-time through gravitational anyonic braids.}
\end{figure}

An initial $|\bar{N}\rangle$ state corresponds to 
a sequence of qu-bits; for example as
\begin{equation}
\label{qquq}
|0,1,0,1,1,...,\rangle\equiv |\alpha, \alpha^{-1}\rangle \otimes |\beta \alpha \beta^{-1},\beta \alpha^{-1}\beta^{-1}\rangle \otimes ... \otimes ...
\end{equation}
The dimensionality of this state is related to the total transition amplitude 
mediating the complete annihilation of all fluxon pairs into nothing. 
This amplitude is interpolated by the many Fusion processes as 
dictated by Eq.\ref{THS}:
\begin{equation}
\label{N}
N_{\alpha\alpha\alpha...\alpha}^{\beta}=\sum_{\{ \beta_{i}\}}N_{\alpha\alpha}^{\beta_{1}}N_{\alpha\beta_{1}}^{\beta_{2}}N_{\alpha\beta_{2}}^{\beta_{3}}...N_{\alpha\beta_{N}}^{\beta}
\end{equation}
$$=\langle \beta|(N_{\alpha})^{N}|\alpha\rangle\, , $$
where $|\alpha\rangle$ corresponds to $|\bar{N}\rangle$ and $|\beta\rangle$ to $|0\rangle$
topological states. 

The $N_{\alpha}$ matrix has eigenstates and eigenvalues
and it can be diagonalized by means of them as
\begin{equation}
\label{N}
N_{\alpha}=|v\rangle D_{\alpha}\langle v|+...+
\end{equation}
where 
\begin{equation}
\label{vav}
|v\rangle =\frac{1}{\mathcal{D}}|D_{\alpha}\rangle,\,\,\,\, |\mathcal{D}|^{2}=\sum_{\alpha}D_{\alpha}^{2}\, .
\end{equation}
The $D_{\alpha}$ are quantum dimensions controlling the single annihilations 
of qu-bites in vacuo. 
Therefore, 
\begin{equation}
\label{NN}
N_{\alpha\alpha\alpha...\alpha}^{\beta}\sim D_{\alpha}^{N}/\mathcal{D}^{2}+...\, .
\end{equation}
This is related to the single annihilation probabilities 
as
\begin{equation}
\label{p}
p(\alpha\bar{\alpha}\rightarrow 0)=(D_{\alpha}/\mathcal{D})^{2}\sim N^{-2}\, ,
\end{equation}
And an annihilation of N pairs correspond to 
\begin{equation}
\label{p}
p(N\, pairs\rightarrow 0)<\, e^{-N}\, ,
\end{equation}
where we used the Stirling approximation,
having included an identical particle factor $N!$.
This provides a topological computing interpretation
of why the CC state with a large $N$ cannot flow to 
the UV nothing state $|0\rangle$. 

\vspace{0.1cm}

{\it The lower dimensional case: punctures and gravitational instantons}. 
It is instructive to consider the 
$2+1$ gravitational case. 
In this contest, we will show a correspondence of gravitational instantons
 with the entropy content 
of de Sitter space-time (see also Ref.\cite{Addazi:2019vch} elaborating on this case).

Let us start considering a Wick rotation of the de Sitter space-time 
as 
\begin{equation}
\label{dSE}
ds_{E}^{2}=\Big(1-\frac{r^{2}}{l^{2}} \Big)dt_{E}^{2}+\Big(1-\frac{r^{2}}{l^{2}} \Big)^{-1}dr^{2}+r^{2}d\theta^{2}\, , 
\end{equation}
where $\beta=2\pi l$ and $0\leq t_{E} \leq \beta$. 
Eq.\ref{dSE} can be rewritten as a metrico of a $S_{3}$ sphere:
\begin{equation}
\label{dSEE}
ds_{E}^{2}=\sin^{2}\rho dt_{E}^{2}+l^{2}d\rho^{2}+l^{2}\cos^{2} \rho\, d\theta^{2}\, , 
\end{equation}
where $r=l\cos \rho$.
The physics of a $2+1$ quantum gravity sector can be captured by 
a double Chern-Simons gauge theory 
with a gauge group $SU_{+}(2)\times SU_{-}(2)$.
Each $\pm$ chiral sector has a connection 
which is a linear combination of the spin connection $\omega^{a}$ and the
triad $e^{a}$:
\begin{equation}
\label{AA}
A_{\pm}^{a}=\omega^{a}\pm \frac{1}{l}e^{a}\, . 
\end{equation}

The de-Sitter background can be recast 
from a class of non-trivial connections
characterized by two real parameters $\gamma,\beta$:
\begin{equation}
\label{e1}
A_{\pm}^{3}=\gamma \cos \rho \Big(d\theta \mp \frac{\beta}{l}dx^{0} \Big)\, , 
\end{equation}
\begin{equation}
\label{e2}
A_{\pm}^{2}=\pm d\rho\, , 
\end{equation}
\begin{equation}
\label{e3}
A_{\pm}^{1}=-\gamma \sin \rho \Big( d\theta \mp \frac{\beta}{l}dx^{0}\Big)\, . 
\end{equation}
The $dS_{3}$ corresponds to the $\gamma=1$ case.
On the other hand, the ratio of the two parameters is 
fixed as a requirement that any no conical singularities of holonomies are present at the horizon
(on $\rho=0$):
\begin{equation}
\label{Aaa}
\beta=2\pi l\gamma^{-1}\, . 
\end{equation}

In the next, we will focus on the $+$ chirality, 
having in mind a specularity of the the negative sector.
The Euclidean action implementing the boundary condition 
reads as
\begin{equation}
\label{IIEE}
S_{E}[A,\beta]=S_{B}+S_{\rho=\pi/2}+S_{\rho=0}\, , 
\end{equation}
\begin{equation}
\label{SB}
S_{B}=\frac{N}{4\pi}\int_{\mathcal{M}}\epsilon^{kl}{\rm Tr}(iA_{k}\partial_{0}A_{l}-A_{0}F_{kl})d^{3}x_{E}\, ,
\end{equation}
\begin{equation}
\label{S1}
S_{\rho=\pi/2}=-\frac{N\beta}{4\pi l}\int_{\rho=\pi/2}\, {\rm Tr}(A_{\phi})^{2}dx^{0}_{E}d\theta,
\end{equation}
\begin{equation}
\label{S2}
S_{\rho=0}=-\frac{N}{2}\int_{\rho=0}A_{\phi}^{(3)}dx^{0}_{E}d\theta\,, 
\end{equation}
compatible with 
\begin{equation}
\label{bou}
A_{0}^{a}|_{\rho=0}=-2\pi \delta_{3}^{a}\, . 
\end{equation}
The euclidean action contains a term on the bulk with a topology $\mathcal{M}=\Sigma \times S_{1}$,
with $S_{1}$ is the compactified euclidean time with periodicity dictated by $\tau=\beta^{-1}$. 
Above, we defined the topological winding number $N$ emerging out as 
\begin{equation}
\label{dee}
N\sim l^{2}/L_{Pl}^{2}\, . 
\end{equation}
The other two terms are on the $\rho=0,\pi/2$ boundaries. 

The corresponding partition function has a saddle semiclassical solution 
on the special class of connections considered: 
\begin{equation}
\label{ZZ}
Z_{A}(\beta)=\frac{1}{\mathcal{N}}\sum_{\gamma}{\rm Exp}\Big(-\frac{1}{16G}\beta \gamma^{2}+\frac{1}{4G}\pi \gamma l \Big)\, ,
\end{equation}
where $\mathcal{N}$ is the normalization factor. 

In other words, the partition function is dominated by the sum 
over $\gamma$, at $\beta$ fixed. 
The saddle point of Eq.\ref{ZZ} 
lies on 
\begin{equation}
\label{gamma}
\gamma \beta=2\pi l\, , 
\end{equation}
which corresponds to the classical $\gamma$-point avoiding for conical singularities 
at the horizon. 
Indeed, the $\gamma$ parametrizes the deficit angle of a conical singularity
located at the extreme point $\rho=\pi/2$ ($r=0$).

Eq.\ref{ZZ} is interpreted as associating 
energy levels and degeneracy factors 
labelled by $\gamma$:
\begin{equation}
\label{Egamma}
E_{\gamma}=\frac{\gamma^{2}}{16G}\, , 
\end{equation}
\begin{equation}
\label{rhoGamma}
\rho(\gamma)={\rm exp}\Big(\frac{\pi \gamma l}{4G}\Big)\, . 
\end{equation}
For $\gamma=1$, the de Sitter entropy is recast as 
\begin{equation}
\label{kjjajaj}
S_=\frac{\pi l}{4G}\sim N_{+}+N_{-}\sim N
\end{equation}
having consider the sum on the $\pm$-chiralities. 
The entropy is related to the degeneracy factor in Eq.\ref{rhoGamma}.

This result can be compared with the semiclassical limit of the 
exact result, computed in terms of spin-network representations 
\begin{equation}
\label{resu}
Z_{A}(\beta)=\sum_{2s}^{N} d^{-1}q^{s(s+1)/N}\frac{\sinh[2\pi(s+1/2)]}{\sinh \pi}f(q,N,s)\, , 
\end{equation}
\begin{equation}
\label{ZAA}
f(q,N,s)=\sum_{n=-\infty}^{+\infty}q^{Nn^{2}+(2s+1)n}e^{2\pi N n}\, , 
\end{equation}
\begin{equation}
\label{dd}
d=\prod_{n=1}^{\infty}(1-q^{n})(1-q^{n}e^{i\theta})(1-q^{n}e^{-i\theta})\, ,
\end{equation}
$$q=e^{-\beta/l},\,\,\,\, \theta=-2\pi i\,. $$

From the semiclassical identification we 
obtain 
\begin{equation}
\label{ide}
n=\frac{\gamma}{2}-\frac{1}{2N}-\frac{s}{N}\, ,
\end{equation}
where $n,s$ are discrete numbers. 
This means that $\gamma$ is quantized. 

This result offers a series of reinterpretations 
as a hint towards a confirmation of the
topological hairon portrait:

\vspace{0.1cm}

(i) the euclidean dS-metric is re-obtained as a the continuous limit of a discrete 
series of conic singularities \footnote{This fact was envisaged in our previous works in Refs. 
\cite{Addazi:2015gna,Addazi:2016cad,Addazi:2015hpa}.}. The conic singularities are saddle solutions related to 
a class of non-trivial connections, i.e. to gravitational instantons. 
In other words, the Euclidean dS Hubble horizon is effectively emerging as a sum on a large number 
of horizonless geometries. 

\vspace{0.1cm}

(ii) The dS-entropy is recast and it is proportional to the instantonic topological number.
The degeneracy factor is exponentially growing as the topological winding number. 
On the other hand, the CC is quantized as the inverse of the topological number. 

(iii) There is a correspondence among spin-network representations and 
gravitational instantons, sustaining the intepretation of the puncture-instanton correspondence.
This can also suggest that the space-time topological computation is related 
to the spin-network dynamics in the bulk in the deep non-perturbative quantum gravity regime. 

It is worth to remark that in the special case of $2+1$ quantum gravity on de Sitter 
all these considerations can be related to the Schwarzian quantum mechanical 
model on the circular time boundary. In turn the Schwarzian model 
is related to the IR conformal limit of the Sachdev-Ye-Kitaev  model (SYK) \cite{Kitaev:2017awl}.
This leads to a dual reinterpretation of hairon fields 
as SYK fields non-locally coupled through a gaussian set of random matrices. 

\vspace{0.2cm}

{\it Entanglement entropy and Topological order}. The topological phase transitions,
the emergence of a topological order
around the space-time quantum criticality, the long-range entanglement,
the exponential degeneracy of the ground state, 
the exotic anyon statistics of punctures, the entropic holographic scaling...All these analogies 
of space-time and 
 topolological materials 
point out towards the definition of a Topological entanglement entropy
of the BH and de Sitter geometries. 
In a topologically ordered two-dimensional disk,
{\it Kitaev, Preskill, Levin and Wen}
found that the von Neuman entanglement entropy
holographically scales 
as 
\begin{equation}
\label{Srho}
S(\rho)=\alpha L-\gamma+...
\end{equation}
where $L$ is the boundary length, in the limit of $L\rightarrow \infty$
\cite{Kitaev:2005dm,TEE2}.
Eq.\ref{Srho} provides a measure of the entanglement of 
the interior and exterior degrees of freedom
related to the density operator as 
\begin{equation}
\label{AA}
S(\rho)=-{\rm tr}\rho\, \log \, \rho\, .  
\end{equation}
The $\alpha$ coefficient depends on particular
aspects of short wave length modes 
close to the boundary. 
On the other hand, $\gamma$ is interpreted as a
universal constant dependent by the feature of the 
ground state entanglement. 
Such a term corresponds to the so dubbed topological entanglement entropy,
depending by the total dimension of the system as 
\begin{equation}
\label{gamma}
\gamma={\rm log}\, \mathcal{D}\sim {\rm log}\, N\, . 
\end{equation}
In our case, we may conjecture that this formula may be generalized to 
a higher dimensional case
as 
\begin{equation}
\label{Srhod}
S(\rho)=\alpha A-\gamma+...\sim N-\log \, N+...\, ,
\end{equation}
by means of the topological order on the boundary area $A$. 
Indeed, in this case, 
the $\gamma$ scales as the 
log of a configuration space $\mathcal{D}$
that is as $N$. 
By means of the thermodynamical definition of entropy,
we arrive to 
\begin{equation}
\label{Srho}
S=-\frac{\partial F}{\partial T}=\frac{\partial}{\partial T}(T\, {\rm log}\, Z)\sim N-\log\, N\, . 
\end{equation}
This relates the topological entanglement entropy to the winding number as 
\begin{equation}
\label{gamma}
N\sim e^{\gamma}\, . 
\end{equation}
The high proliferation of N states 
is connected to a positive topological entanglement entropy. 
This is another way to visualize that 
a changing of BH entropy corresponds to an
alteration of the topological order in vacuo.

\vspace{0.4cm}

{\it Final remarks}. In this paper, we explored the connections among
the $\mathcal{HN}$ paradigm and gravitational topological properties of the vacuum state.
We showed that quantum hairs stored in space-time are intimately related to the 
topological winding number. The CC corresponds to a maximal entropic vacuum state, 
where the information storage is critically enhanced. Within this picture, the de-Sitter space-time 
lives in a critical conformal phase, where the critical temperature is exactly related to the CC. 
Around the critical temperature, there is a Topological Phase Transition of the space-time geometry. 
Indeed, the CC is stabilized by the maximal entropic and topological content of the Universe vacuum state:
a IR to UV destabilization is exponentially suppressed. 
Then, we elaborated on the topological meaning of hairon fields. 
We realized that they can be interpreted as punctures on the space-time boundary 
from gravitational Wilson lines or Wormholes crossing the dS-bulk. 
The hairon spin statistics is neither fermionic nor bosonic:
hairons are gravitational non-abelian anyons. This is caused by the topological complexity of the bulk as a Wilson line network. 
Such a topological picture is potentially insightful for our understanding of the space-time and the Black Hole
information storage. 
It explains why the space-time has an exponentially enhanced memory and a area law, symptomatically suggesting that 
a large number of quantum hairs are in a degenerate entangled ground state. 
As in topological quantum computers, the effect of any fluctuations or noise, potentially creating or destroying any new qu-bits, 
is exponentially suppressed. This offers a quantum information portrait of the CC stabilization mechanism:
the gravitational vacuum topology provides a cosmological defense of the CC memory. 
Any CC-Planck scale mixings, corresponding to a massive attack to the Universe Memory, are highly suppressed
as a non-local topological protection of the hairon/anyon entangled state. 


\onecolumngrid


\twocolumngrid

\end{document}